\documentclass[10pt,conference]{IEEEtran}

\usepackage{amsmath}
\usepackage{graphicx}
\usepackage{amssymb}
\usepackage{amsthm}
\usepackage{verbatim}
\usepackage{epsfig}
\usepackage{color}
\usepackage{url}
\usepackage{listings}
\usepackage{subfigure}
\usepackage{stmaryrd}
\usepackage{siunitx}
\usepackage{siunitx}

\newcommand{\bugnum}[1]{Bug-#1}
\newcommand{\code}[1]{\textsf{\small #1}}

\ifx\usepdfsync\undefined
\else
  \usepackage{pdfsync}
\fi

\begin{document}

\title{Debugging Behaviour of Embedded-Software Developers:~An Exploratory Study}

\author{\IEEEauthorblockN{Pansy~Arafa,~Daniel~Solomon,~ Samaneh~Navabpour,~Sebastian~Fischmeister} 
\IEEEauthorblockA{
Dept. of Electrical and Computer Engineering\\
University of Waterloo, Canada\\}
\{{\normalfont parafa, d4solomo, snavabpo, sfischme\}@uwaterloo.ca}}

\maketitle

\begin{abstract}

Many researchers have studied the behaviour of successful developers while debugging desktop software.
In this paper, we investigate the embedded-software debugging by intermediate programmers through an exploratory study.
The bugs are semantic low-level errors, and the participants are students who completed a real-time operating systems course in addition to five other programming courses.
We compare between the behaviour involved in successful debugging attempts versus unsuccessful ones.
We describe some characteristics of smooth and successful debugging behaviour.
\end{abstract}

\section{Introduction}
\label{sec:introduction}

Embedded systems represent 98\% of all computers (DARPA, 2000).
An embedded system must accurately fulfill its functional requirements in addition to strict timing and reliability constraints.
Examples of embedded systems are high-performance networks and robotic controllers. 
More applications of embedded systems are safety critical. 
They include the new generation of airplanes and spacecraft avionics, and the braking controller in automobiles.
The correctness of these systems depends, not only on the results they produce, but also on the time at which these results are produced~\cite{book1}.
The verification process of such systems is complicated, and consequently many testing techniques and tools exist to ensure system's accuracy.
That intensifies the necessity of enhancing developer's debugging skills.

As an essential part of software-system development, debugging is a difficult and expensive task.
A study in 2002 has revealed that software engineers spend an average of 70\%-80\% of their time testing and debugging~\cite{Tassey2002}.
We can define debugging roughly as the process of error detection and maintenance of functional correctness.
For desktop applications, many researchers conducted experiments to determine the behaviour which promotes successful bug repair and time management.

In this paper, we focus on the characteristics of successful debugging of embedded software.
Debugging of embedded software is challenging due to 
(1) hardware interaction, e.g., loading code to the target board, 
(2) use of low-level language semantics, e.g., memory management in C, 
and (3) the need to respect the system's timing requirements.
We believe the enhancement of the developers' debugging skills makes the current research work in the embedded-systems domain more promising and beneficial for the industrial market.
That research work includes testing of embedded automotive communication systems~\cite{Armengaud2008},
time-aware instrumentation of embedded software~\cite{Arafa2013, Kashif2013}, 
 and debugging of concurrent systems~\cite{Musuvathi2008}.

In this paper, we present an exploratory study, in which 14 intermediate-level programmers debug semantic errors of embedded software.
In the study, we use seven distinct bugs classified into two categories: incorrect hardware configuration and memory leaks.
We explore the debugging behaviour of the participants and provide comparisons between successful and unsuccessful attempts.
Moreover, we introduce the activity visitation pattern: a graphical representation for the debugging behaviour.


 \vspace{-10pt}
\section{Related Work}
\label{sec:relatedwork}
\vspace{-4pt}
Many program-comprehension studies, since the 1970s, have inspected software debugging techniques of developers.
As one of the first, Gould examined the behaviour of experienced programmers while debugging non-syntactic errors in FORTRAN programs~\cite{Gould1975}. 
One goal of such studies is understanding the activities associated with the debugging and testing processes~\cite{Mayrhauser1997, Karahas2007, Murphy2008, fitzgerald2010}.
For instance, Murphy et al.~\cite{Murphy2008} presented a qualitative analysis of debugging strategies of novice Java programmers.
The authors stated twelve categories of debugging strategies used by the subjects and discussed their effectiveness, e.g., tracing, pattern matching, and using resources.
Another objective is the identification of the difficulties faced by the developers during software-change or maintenance tasks.
Examples of debugging challenges include  bug localization~\cite{fitzgerald2010}, understanding unfamiliar code~\cite{fitzgerald2010}, and gaining a broad understanding of the system~\cite{Sillito2005}.
Moreover, there exists a wide range of empirical and exploratory studies that examined the debugging techniques of different experience levels of programmers, and also, compared successful to unsuccessful behaviour~\cite{Robillard2004, Sillito2008, Murphy2010, Mayrhauser1996}.
Lastly, implications for teaching and evaluation of the existing development tools based on the studies' insights are presented in~\cite{Robillard2004, Sillito2005, Ricca2007, Murphy2008}, and~\cite{Sillito2008}.
To our knowledge, this paper is the first work to address embedded-software debugging, which includes low-level programming and hardware interaction.
\section{Methodology}
\label{sec:methodology}

\subsection{Participants}
\label{sec:participants}
The study's participants are intermediate-level developers. Each of them met the following criteria:
\begin{itemize}
\item They have completed a \emph{Real-time Operating Systems} course in the previous year in addition to five different programming courses throughout their undergraduate studies.
\item In the \emph{Real-time Operating Systems} course, they implemented a real-time operating system (RTOS) on a Janus MCF5307 ColdFire Board similar to the study's platform.
They also have worked on the same RTOS in another course.
\item They have had an average of four co-operative work terms (about 2400 working hours in total).
\end{itemize}
Participants meeting these criteria have sound knowledge of the board and the type of program they are going to debug.
Hence, these participants have \emph{intermediate} programming skills for embedded-software development.
We scheduled each participant in a separate two-hour time slot.

\subsection{System}
\label{sec:system}
The participants attempted to fix bugs contained within a small real-time operating system (RTOS) implemented on a Janus MCF5307 based microcontroller board.
The RTOS is in C language; it consists of 23 C files and 15 header files.
In total, it has 3085 lines of code (LOC).
That RTOS is similar to the one they have implemented in the previous year.
So the participants in our study are \emph{familiar} with such \emph{low-level} software and have reasonable experience with the RTOS.

The RTOS supports a basic multiprogramming environment with five priority levels and preemption.
It also supports simple memory management, inter-process communication, basic timing services, an I/O system console, and debugging support.
The RTOS contains a fully integrated timer, a dual UART (universal asynchronous receiver/transmitter), and several other peripheral interface devices.
The RTOS and the application processes use up to one megabyte of RAM.
Software development for the RTOS is supported by gcc and the ColdFire/68K cross-compiler.
During the study sessions, we provided the participants with the RTOS documents and the Janus MCF5307 ColdFire Board manual.

\subsection{Bugs}
\label{sec:bugs}
This study includes seven semantic bugs categorized into incorrect-hardware-configuration bugs and memory leaks.
Misconfiguration and memory errors constitute a notable share of the root causes of software bugs~\cite{Li2006, Yin2010}.
Each bug is provided with a report that states the buggy behaviour of the system.
The participant examines only one bug at a time.

The first  bug category is incorrect hardware configuration.
\mbox{\bugnum{1}} report states the following: "Interaction with the OS does not result in understandable messages; corrupted text appears on the screen."
\mbox{\bugnum{1}} is an incorrect assignment of the serial port baud rate (Line~\ref{bug1loc} in Listing~\ref{Bug1Code}).
Thus, the system shows corrupted text on the screen when communicating.
\mbox{\bugnum{2}} has an incorrect value of timer's match register.
This causes the timer interval to be three times slower than intended.
\mbox{\bugnum{7}} causes the board to reset about six seconds after launching the OS because the function \code{resetWatchdog()} resets the Watchdog timer to a wrong value.
\begin{minipage}[htb]{.9\linewidth}
\lstset{basicstyle=\small, numbers=left, numberstyle=\tiny, firstnumber=50, stepnumber=1, numbersep=4pt}
\begin {lstlisting} [language=c, breaklines=true, frame=lines, label={Bug1Code}, caption={\bugnum{1} code},captionpos=b,aboveskip=12pt, escapeinside={@}{@}]
int initUARTIProc(){
    ...
    /* Set the baud rate (19200) */
    SERIAL1_UBG1 = 0x00;
    @\label{bug1loc}@SERIAL1_UBG2 = 0x49; 
    /* Set clock mode */
    SERIAL1_UCSR = 0xDD;
     ...
}
\end{lstlisting}
\end{minipage}

The second-category bugs contain memory leaks.
In \mbox{\bugnum{3}}, the OS stops responding shortly after launching one of the user processes because it fails to release messages that it receives.
This causes the OS to run out of user memory after two iterations of the process.
Listing~\ref{Bug3Code} shows the code portion of  \mbox{\bugnum{3}}.
Two statements are missing: \code{release\_memory\_block(pMsgDelay)} \emph{directly} after Line~\ref{bug3aloc}, and \code{release\_memory\_block(pMsgIn)} \emph{directly} before Line~\ref{bug3bloc}.
In  \mbox{\bugnum{4}}, there is a missing function call in an interrupt process to release messages that it prints to the terminal.
This causes the OS to run out of user memory after some interaction.
A similar problem occurs in  \mbox{\bugnum{5}}; the Wallclock process fails to release the delay messages.
As a result, the OS stops responding some time after launching the Wallclock process.
In  \mbox{\bugnum{6}}, although the proper \code{release\_memory\_block()} calls exist in the OS, no memory is actually freed since the boundary checks of the user memory address range are incorrect.

\begin{minipage}[htb]{.9\linewidth}
\lstset{basicstyle=\small, numbers=left, numberstyle=\tiny, firstnumber=100, stepnumber=1, numbersep=4pt, tabsize=1}
\begin {lstlisting} [language=c, breaklines=true, frame=lines, label={Bug3Code}, caption={\bugnum{3} code}, captionpos=b, aboveskip=12pt,belowskip=6pt, escapeinside={@}{@}]
void ProcC(){
	...
	while(1){
		...		
		if((pMsgIn->pData)[0] == 0){
			pMsgOut = request_memory_block();	
			...
			send_message(PID,pMsgOut);	
			while(1){
				pMsgDelay = receive_message();		
				if(pMsgDelay->msg_type == wakeup){@\label{bug3aloc}@						
						break;
				}else{...}
			}
		}		
		release_processor();@\label{bug3bloc}@
	}
}
\end{lstlisting}
\end{minipage}

\subsection{Data Collection}

Our exploratory study involved three methods for data collection:
\begin{enumerate}
 \item We video-recorded the computer screen during the study sessions using CamStudio~\cite{Cam}.
 Three observers coded the videos using the CowLog software~\cite{Cowlog} to mine the videos and extract information revealing the debugging behaviour.
For each bug per participant, the video-mining process resulted in time-stamped files that show (1) the examined bug, (2) the activity, e.g., code browsing and editing,
(3) the code module, and (4) the debugging technique such as tracing and adding print statements.
 \item Each time a participant compiles the system, the system is copied to a \emph{.try} folder. 
 The \emph{.try} folders are used later to view the edits made in every compilation try.
 \item Each participant filled out a pre-session and a post-session form in addition to a post-bug form for each bug he examined.
\end{enumerate}

\section{Results and Discussion}
\label{sec:resultsanddiscussion}

\newcounter{ObsID} 

This section lists our observations based on the collected data. These observations (1) highlight the debugging activities used by the developers to fix the low-level bugs, and (2) compare between the successful debugging attempts versus the unsuccessful ones.
Table~\ref{tab:bugsummary} presents a summary for the seven bugs in the study.

{\bf Definitions.} 
The total examining time of a bug is considered to be the whole time spent by the participant to investigate that bug (including all debugging activities).
The total editing time refers to the entire time spent editing the program, which includes insertion, modification, and deletion of code lines.
We define the time spent to locate a bug as the time elapsed before the first edit to the right function.
Additionally, we define the time to fix a bug as the difference between the total examining time and the time to locate the bug, which is calculated only for bugs that have been fixed. Finally, we consider a bug is located if the participant edited the right function.

\begin{table*}
\begin{center}
 \caption{Bug Summary}
 \label{tab:bugsummary}
  \begin{tabular}{| c | l | c | c | c | c | c |} \hline
    \bf{ID} & \bf{Name} & \bf{Category} & \bf{Examined By} & \bf{Located By} & \bf{Fixed By} & \bf{Average Examining Time}  \\ \hline
    1 & Incorrect Baud Rate & Incorrect HW Config. & 14 & 36\% & 21\% & 46 min  \\ \hline
    2 & Incorrect Timer Register Value & Incorrect HW Config. & 7 & 43\% & 14\% & 30 min   \\ \hline
    3 & Memory Leak in User Process C & Memory Leak & 11 & 91\% & 36\% & 37 min \\ \hline
    4 & Memory Leak in UART & Memory Leak & 6 & 67\%  & 50\% & 23 min \\ \hline
    5 & Memory Leak in Wallclock & Memory Leak  & 4 & 100\% & 100\% & 15 min \\ \hline
    6 & Incorrect Memory Boundary check & Memory Leak  & 3 & 33\% & 33\% & 26 min \\ \hline
    7 & Incorrect Watchdog Value & Incorrect HW Config. & 2 & 100\% & 100\% & 17 min\\ \hline
  \end{tabular}
\end{center}
\vspace*{-6mm}
\end{table*}





\stepcounter{ObsID}
{\bf Observation \theObsID.}
Locating the bug is harder than fixing it.
In 93.75\% of the post-bug forms filled out by the participants,
they considered the difficulty of locating the bug to be higher than or equal to the difficulty of fixing the bug (30 out of 32).
In all but one cases, the time spent to locate the bug is higher than the time spent to fix it.
In total, the participants located only 63.82\% of the bugs (30 out of 47).

\begin{figure}[htb]
\centering
\subfigure[Successful attempts.]{
\includegraphics[width=.6\linewidth, height=.6 \linewidth]{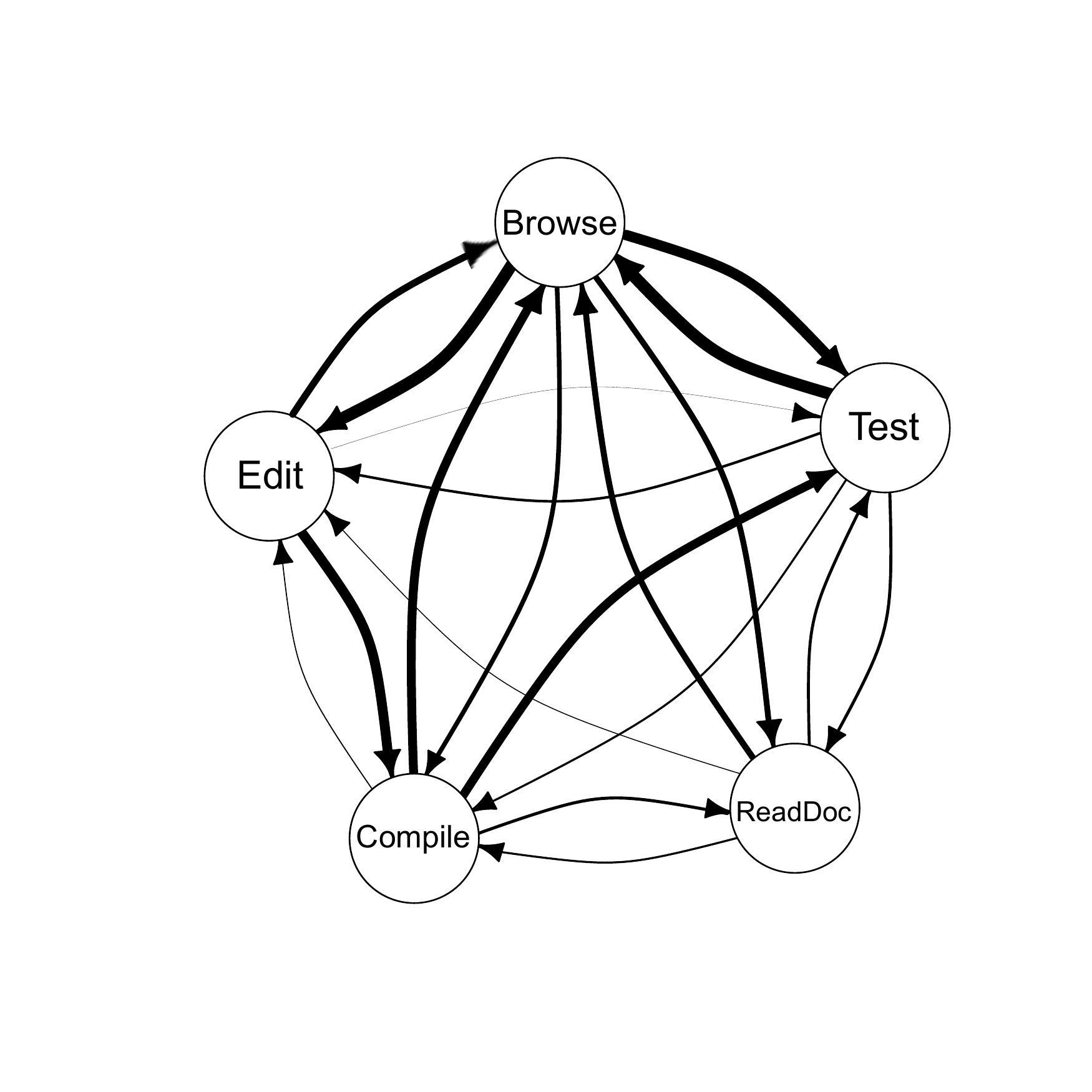}
\label{fig:visit1}
}
\subfigure[Unsuccessful attempts.]{
\includegraphics[width=.6\linewidth, height=.6 \linewidth]{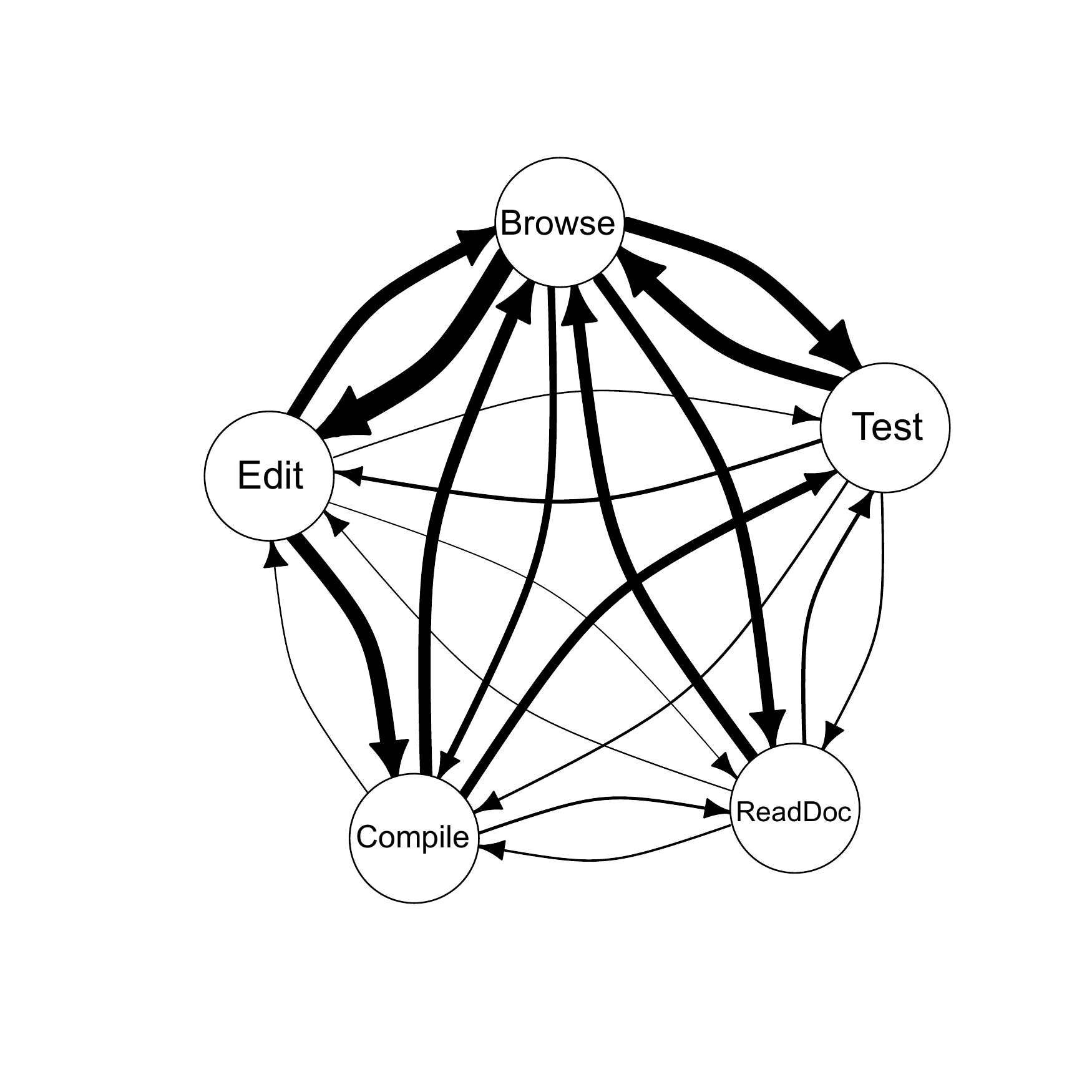}
\label{fig:visit2}
}
\label{fig:visitpattern}
\caption{Activity visitation patterns}
\end{figure}

\stepcounter{ObsID}
{\bf Observation \theObsID.}
We investigated the \emph{activity visitation pattern}, which shows the frequency of transition from each activity to another while examining a bug.
The activities are code browsing, code editing, document reading, compiling, and testing.
Figure~\ref{fig:visit1} shows the activity visitation pattern of the successful debugging attempts, i.e., the cases were the participants successfully fixed the bug.
Figure~\ref{fig:visit2} presents the pattern of the unsuccessful attempts where the participant failed to fix the bug.
Every node represents a debugging activity.
We calculated the average value of the number of transitions between each two activities, and accordingly, set the thickness of the transitions.
Thus, the thickness of a transition represents the frequency of switching from the activity of the source node to that of the destination node.

In general, the unsuccessful attempts have a higher number of transitions than the successful ones.
The average number of transitions between activities are 43.11 and 26.97 in the unsuccessful and successful attempts consequently.
We believe that the high number of transitions in the unsuccessful attempts conveys the indecisive behaviour of the participants which can be a reason for failing to fix the bugs.

Figure~\ref{fig:visit1} demonstrates the path of code browsing followed by code editing then compiling and finally testing.
This sequence repeats until the bug is fixed. Although that path exists in Figure~\ref{fig:visit2}, many other transitions interpose it.
We believe that the mentioned path supports a smooth and successful debugging process.
In Figure~\ref{fig:visit1}, the transition from the code browsing activity to the testing one is frequent.
This may convey the attempts of the participants to comprehend the program based on its output.



\stepcounter{ObsID}
{\bf Observation \theObsID.}
Successful debugging behaviour \emph{considers alternatives}, which means considering that a bug can have multiple causes.
In \mbox{\bugnum{3}}, two lines of code are missing, although, many participants inserted only one line.
Eleven participants examined this bug and 10 located it.
However, only three participants successfully fixed it.

\stepcounter{ObsID}
{\bf Observation \theObsID.}
Successful debugging behaviour implies editing only one function in a compilation try.
This is equivalent to avoiding \emph{multiple edits} behaviour.
The absence of multiple edits in all tries correlates with low total editing time, low time to locate bug, and low total examining time. 
However, the presence of multiple edits in any try does not necessarily correlate with high total editing time, high time to locate bug, etc.
The multiple edits behaviour was avoided in 61.11\% of the successful debugging attempts (11 out of 18).

\stepcounter{ObsID}
{\bf Observation \theObsID.}
Successful debugging behaviour requires running the first compilation try without any changes to the program to investigate the behaviour of the buggy system.  
88.88\% of the successful attempts maintained this behaviour (16 out of 18).  
86.66\% of the located-bug cases maintained this behaviour (26 out of 30).  
75\% of the participants who edited the program before the first compilation try failed to fix the bug (6 out of 8).

\begin{figure}[htb]
\vspace{-3mm}
\centering
\includegraphics[width=.9\linewidth]{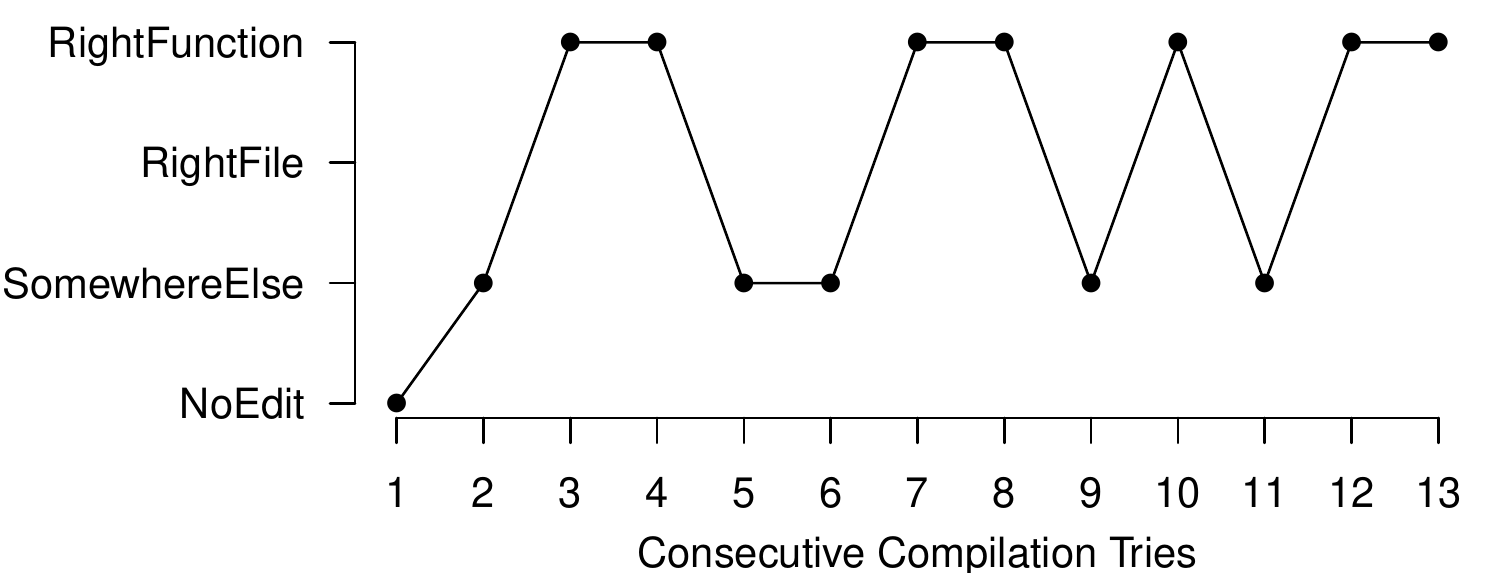}
\vspace{-2mm}
\caption{Example of the ping-pong behaviour (Participant-7,  \mbox{\bugnum{3}})}
\label{fig:ping}
\vspace{-3mm}
\end{figure}

\stepcounter{ObsID}
{\bf Observation \theObsID.}
Successful debugging behaviour maintains a smooth approach to locating and fixing bugs with almost no retesting for a previously tested function. 
That is equivalent to avoiding \emph{ping-pong behaviour}. It means, according to the editing location, moving far from the bug after approaching it at least twice.
Figure~\ref{fig:ping} shows an example of the ping-pong behaviour in Participant-7 debugging session of  \mbox{\bugnum{3}}. The x-axis represents the number of the compilation try. The y-axis represents the location in the program that the participant edited in each try. 
There are four editing categories:
\begin{enumerate}
\item The participant edits the right function which contains the bug. That means the participant located the bug.
\item In the right file which includes the bug, the participant edits any function except the right one.
\item The participant edits somewhere other than the file containing the bug.
\item The participant compiles the system without making any modifications. 
\end{enumerate}
Figure~\ref{fig:ping} shows a non-smooth debugging process.
The participant made no code changes in the first compilation try.
In the second try, the participant edited a function somewhere other than the right file.
Then, he modified the right function in the third and the fourth tries, which means he located the bug;
but in the fifth and the sixth tries, he again edited a function far from the bug location.
He switched between editing the right function and editing somewhere else four more times.
It looks like random edits in different functions, files, and code modules.
The absence of the ping-pong behaviour correlates with low time spent to locate bug and low total examining time,
but the presence of the ping-pong behaviour does not necessarily correlate with high time spent to locate bug or high total examining time.
88.88\% of the successful debugging attempts did not include the ping-pong behaviour (16 out of 18).
Also, 80\% of the located bug cases did not include the ping-pong behaviour (24 out of 30),
and 75\% of the ping-pong behaviour cases are unfixed (6 out of 8).

\section{Threats to Validity}
\label{sec:validation}

{\bf Video Coding.}
The process of analyzing the videos is challenging due to the difficulty of categorizing human behaviour.
Three observers coded the videos such that the videos of six participants were coded by two observers in conjunction.
To validate the information extracted from the videos, we calculated the inter-rater reliability value using Fleiss' kappa method for categorical ratings.
It is equal to 0.74 which indicates substantial agreement among the three observers.

{\bf Learning Effect.} The bugs' identifiers did not introduce biased results.
While the bugs are numbered in the same order, the participants were free to choose the bugs to inspect.
They examined different sets of bugs (e.g., one participant examined bugs 1, 3, 5 and 6; another one examined bugs 1, 2, 3 and 4).
We found no correlation between the examining time and the bug identifier.
Also, there is no correlation between the examining time and the order in which each participant examined the bugs.

{\bf Bug Localization.} We consider a bug is located if the participant edited the right function, but in some cases, he might not recognize that he located it (e.g., inserting a print statement).
Since we can not read the participant's mind, we had to set such an assumption to provide an accurate measurement for locating bugs.
On the other hand, it is a reasonable assumption since all the bug-containing functions are short; they have an average of 37 LOC.

\section{Conclusion}
\label{sec:conclusion}

In this paper, we presented an exploratory study of the debugging behaviour of intermediate embedded-software developers.
We also demonstrated the use of the activity visitation pattern.
In general, debugging is a complex and time-consuming task especially for embedded software.
Understanding the debugging behaviour of developers should be achieved before the creation of aiding tools.
This work provided results that can guide future researchers on automated debugging tools for the embedded domain. 




\bibliographystyle{IEEEtran}
\bibliography{refs}

\end{document}